\documentclass[10pt,conference]{IEEEtran}
\pdfoutput=1
\usepackage[pdftex]{graphicx}
\usepackage{amsmath}
\usepackage{multirow}
\usepackage{balance}
\DeclareRobustCommand*{\IEEEauthorrefmark}[1]{\raisebox{0pt}[0pt][0pt]{\textsuperscript{\footnotesize #1}}}
\begin{document}
\title{Integrating Queuing Regime into Cognitive Radio Channel Aggregation Policies: A Performance Evaluation}
\author{\IEEEauthorblockN{
		Ebenezer Esenogho\IEEEauthorrefmark{1},   
		Elie N. Mambou\IEEEauthorrefmark{2},   
	} 
	\IEEEauthorblockA{\IEEEauthorrefmark{1}
		Centre for Telecommunication, Dept. of Electrical and Electronic Engineering Science, University of Johannesburg, \\P. O. Box 524, Auckland Park, 2006, South Africa, ebenezere@uj.ac.za}
	\IEEEauthorblockA{\IEEEauthorrefmark{2}
		Centre for Telecommunication, Dept. of Electrical and Electronic Engineering Science, University of Johannesburg, \\P. O. Box 524, Auckland Park, 2006, South Africa, emambou@uj.ac.za}
}

\maketitle

\begin{abstract}
Channel aggregation (CA) is one of the newest concept which cognitive radio network is bringing to bear for the smooth role out of fifth/next generation wireless networks. This is the combining of several unused primary user spectrum holes into a logic usable channel. However, several of these strategies have been investigated considering the varying nature of wireless link and adaptive modulation and coding (AMC). Examples are the instant blocking strategy (IBS) and readjustment based strategy (RBS). This paper develops and compares two CA policies with queue, which are the IBS with queue (IBS + Q), and the RBS with queue (RBS+Q). This is in furtherance of previous proposed work. The aim is to identifying the impact of a queuing regime on the performance of the secondary network such that any secondary user (SU) that has not completed its service, as an alternative to dropping or forcibly terminating the service, it is queued in order to get another opportunity to access the primary user (PU) channels. The performance is evaluated through a simulation framework. The results validate that with a well-designed queuing regime, capacity, access and other metrics can be improved with significant reduction in blocking and forced termination probabilities respectively.
\end{abstract}

\vspace{7pt}
\section{Introduction}\label{sec1}
Over the year research has shown that cognitive radio is a promising model for supporting dynamic and opportunistic spectrum access \cite{liang2011, domenico2012}. This is based on its capability to dynamically and autonomously alter its functioning constraints (transmission parameters), learn from surroundings and make decisions based its previous experience \cite{nguyen2012, su2008}. However, the erratic usage of spectrum by PU creates spectrum holes. Therefore, this scattered spectrum holes need organisation to meet the SU experience requirement and as such, the need for CA policy. CA policy allows SU (unlicensed user) to combine several free spectrum holes to improve SU throughput and same time reduce PU blocking and forced termination of SU service \cite{jiao2012}. Furthermore, the capabilities of cognitive radio can be categorized under into spectrum sensing, spectrum management, spectrum adjustment and spectrum sharing etc. This paper will be focus on CA as a type of spectrum sharing, which entails the co-ordination between SUs to access the channel co-operatively or otherwise.
CA can not be discussed in isolation form spectrum adaptation. As earlier mentioned, the introduction of a queuing regime for SUs is to enhance traffic flows so that service that would have been blocked or forcibly terminated could then be queued in a buffer and possibly served later. However, CA can be designed and implemented with or without queues. Several of which without queue were proposed in \cite{jiao2013}; while \cite{balapuwaduge2014} and \cite{esenogho2015} considered CA integrating queuing regime. This paper, focused precisely on work done in \cite{esenogho2015}, \cite{esenogho2015b} which proposed the instant blocking strategy $(IBS)$ and readjustment based strategy $(RBS)$. In addition, this paper incorporates a two feedback mechanisms for SUs which was not considered in previous studies.

The policy though considered the dynamics of the wireless link with $AMC$, but did not take into account joint policy that integrate queueing regime into the $IBS$ and $RBS$ schemes such that, services that would have been blocked or dropped are queued and served later depending on the wireless link condition.

The rest of the paper is organised as follows: Section II summarized related work. The system model of the proposed strategy is presented in Section III. Aggregation policies and system model performance measures are captured in Section IV and V respectively. Numerical results with corresponding discussions are illustrated in Section VI. And finally, this paper is concluded in Section VII. 
\vspace{7pt}
\section{Related work}\label{sec2}
In \cite{ngatched2013}, the performance analysis of two CA policies with imperfect detection for wideband CRN were proposed. In \cite{li2012}, policies integrate queuing and channel fragmentation in channel assembling. This is a novel approach though, the scope of this work is not on fragmentation strategy.  \cite{jiao2012} considered twofold spectrum adjustment strategies in a heterogeneous traffic scenarios. The performance of these policies is evaluated and analysed on a proposed continuous time Markov model. \cite{jiao2013} developed and investigated a mathematical equation for the hypothetical upper limit of the SU network with channel assembling. 
In \cite{jiao2011}, CA for real time traffic with channel adjustment were studied subject to channel accessibility and other SU events. In \cite{jiao2010}, the evaluation of channel assembling was investigated when spectrum adjustment was not applied. In the study, scenarios that was taken into account are without CA, with static and dynamic assembling, when SUs are granted admission into the network. Closely related works \cite{esenogho2015b} proposed and compares through an investigative study, two CA policies. In the study, SUs services are instantly blocked if there are inadequate spectrum resources; while in the $RBS$, the spectrum is adjusted to accommodate new arrivals or other users. However, a queueing regime is integrated into the $IBS$ and $RBS$ schemes to proposed $IBS+Q$ and $RBS+Q$ such that those services that would have been blocked or dropped are queued and served later using two feedback mechanisms.   
\vspace{7pt}
\section{System model}\label{sec3}
\begin{figure}[h]
	\centering
	\includegraphics[width=1\linewidth]{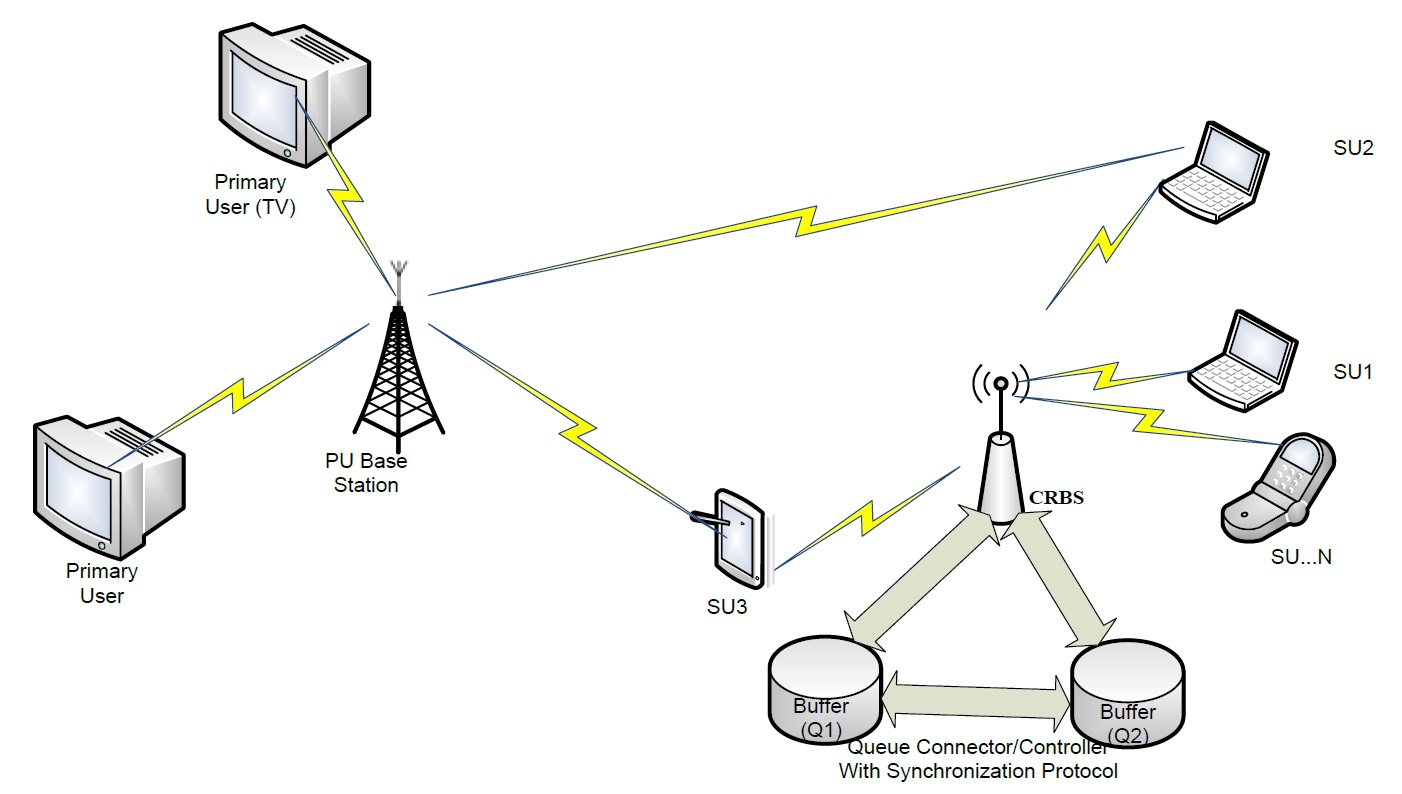}
	\caption{Network Architecture.}
	\label{fig:fg1}
\end{figure}

The centralized network architecture comprises of two independent controllers called the primary user base station $(PUBS)$ and cognitive radio base station $(CRBS)$ respectively. These controllers have respective SUs, using the same spectrum as illustrated in Fig. \ref{fig:fg1}. 

\begin{figure}[h]
	\centering
	\includegraphics[width=1\linewidth]{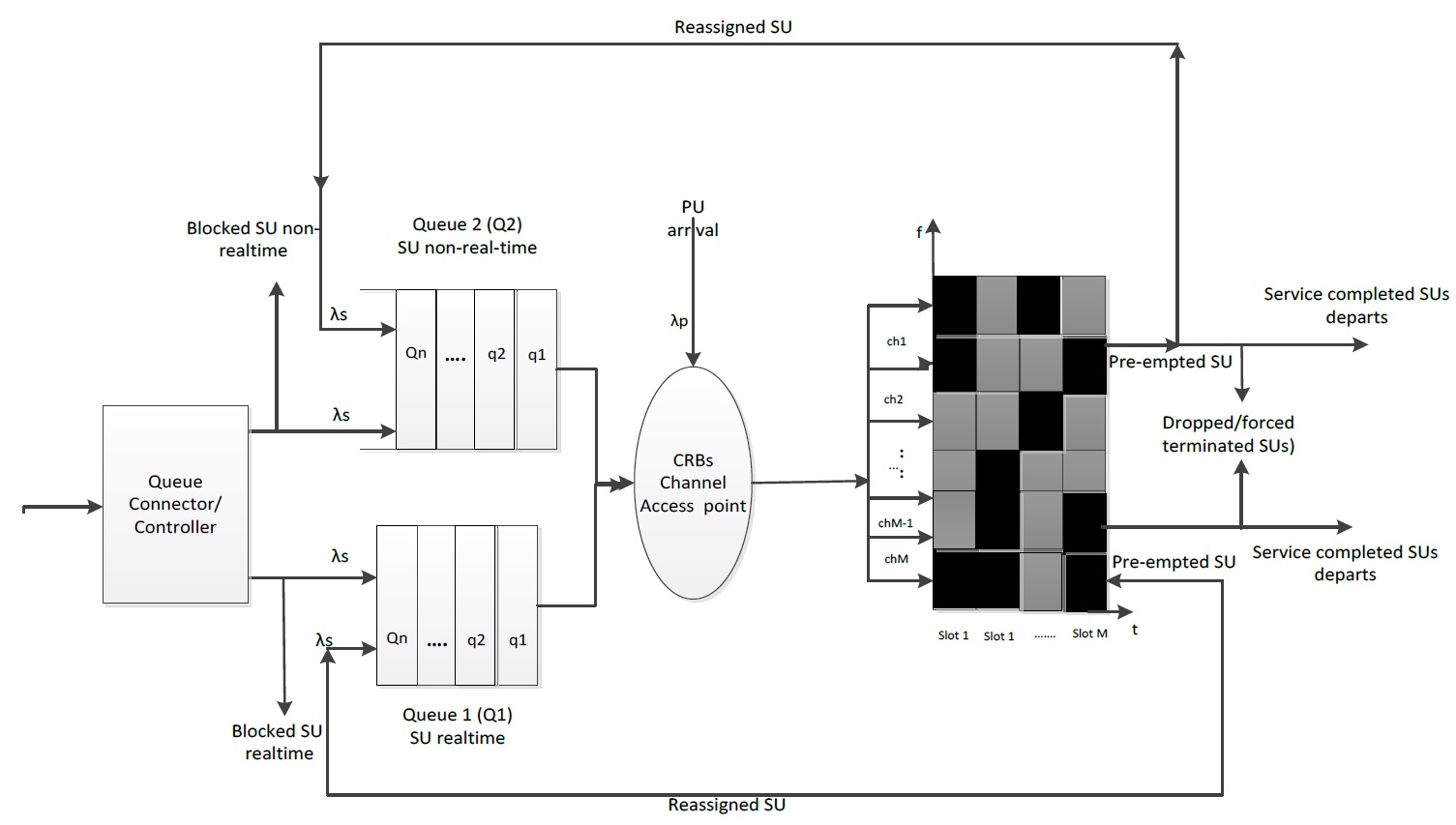}
	\caption{Schematic of the proposed queuing regime deployed.}
	\label{fig:fg2}
\end{figure}

The $CRBS$ have two queuing regimes synchronised to compensate each other in the event of overflow due to batch arrival of a class of SUs. The queuing controller select SUs with a first in first out (FIFO) protocol. The PU needs a channel slot while a SU combines more than one channel using the orthogonal access scheme, as shown in Fig. 3. The PU behaviour is characterized as a busy/idle while the SU optimistically and resourcefully aggregates several neighbouring channels. The essence of the two feedback flow in Fig. \ref{fig:fg2}, is to enable SUs get the opportunity to access the spectrum after its service has been forcibly terminated. Forced termination/dropping occurs in two folds: a) When the PU arrives and no other available channel for the SU to switch to; b) Within the queue when the SU over stayed the queue. However, the later occurs at worse case scenarios. 

The channel utilization chart for a PU busy$(ON)$/idle$(OFF)$ activities is illustrated in Fig. 3. The channel state utilization of PU is characterized as Markovian process as in \cite{esenogho2015, esenogho2015b}, with $C_i$, being the transition probability from $ON$ state to $OFF$ state and $A_i$ being the transition likelihood from the idle state to the $ON$ state for the  $l^{th}$ channel. For SUs, the CRBS detects diverse SNR since all SUs would not be the same. So, we assume the SU SNR will falls under good, moderate or bad SNR. The consequence of this flexibility is a heterogeneous structure with variable resource capacity such that SU traffic requires a precise communication rate or number of channel slots for a certain channel link condition \cite{esenogho2015, esenogho2015b}. 

\begin{figure}[h]
	\centering
	\includegraphics[width=1\linewidth]{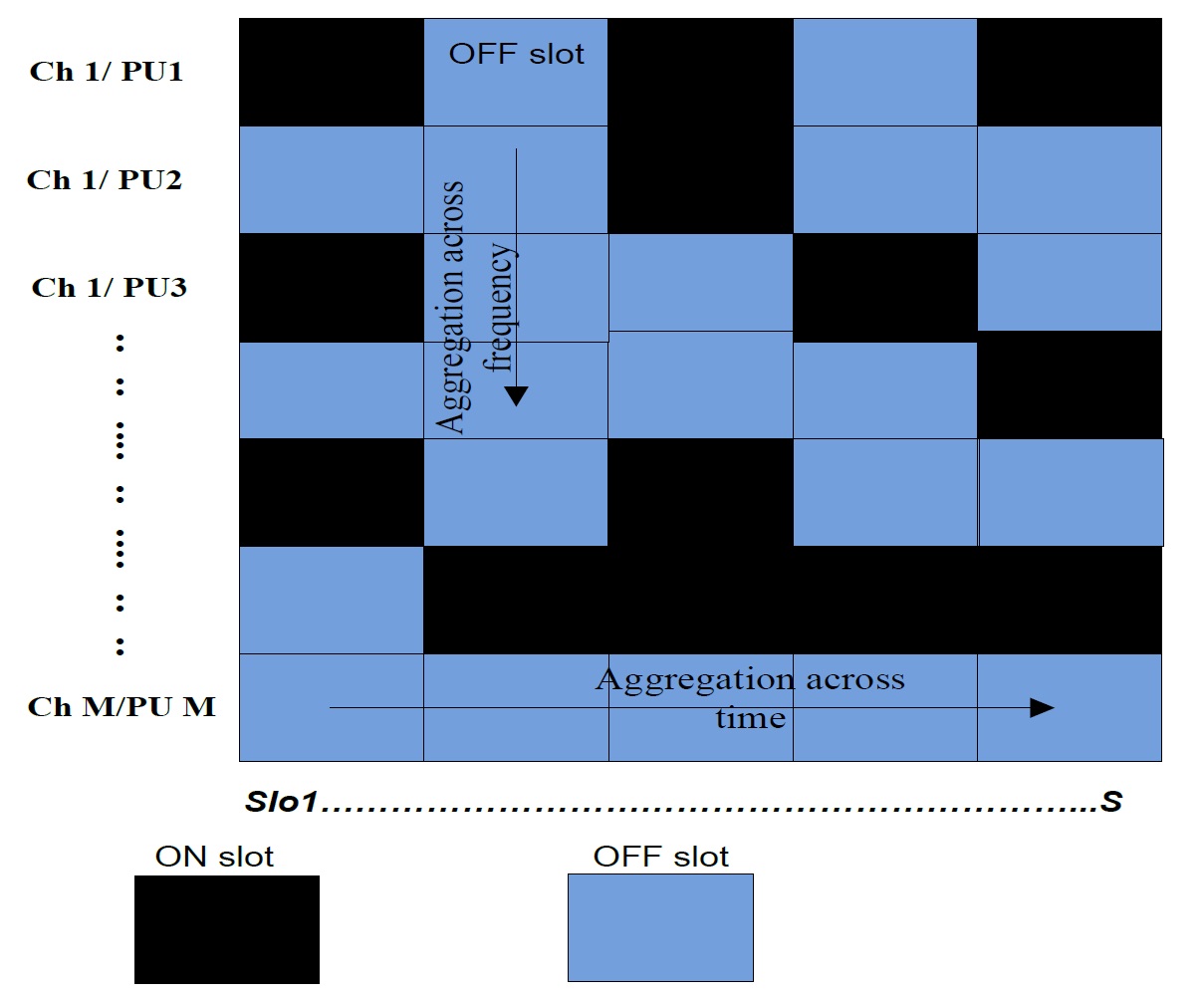}
	\caption{SU frame utilizing OFDMA.}
	\label{fig:fg3}
\end{figure}
\subsection{Link Model and AMC}
$Nakagami-m$ channel model is used to describe the wireless channel owing to its flexibility of covering bound of dwindling channels \cite{wang2007}. The channel quality is captured by the SNR while varying conditions are characterized by the Markov model whose analysis for slow fading channel conditions are well established in \cite{wang2007}. 
\subsection{AMC and SU Frame configuration}
Let the message length in bits be denoted as $\pi$ and  $r_n$ remains the number of bits carried per symbol for mode $n$, where the mode ranges $1 \leq n \leq N$, and $N$ being the highest mode. The number of channel in a static frame $S$ is a variable depending on a varying SNR denoted as $\gamma$. $S$ also corresponds to the number of slots in a frame for a coherent time interval. It can be expressed as: 
\begin{equation}
S=\Big(\frac{\pi}{r_n\varepsilon_sr_s}\Big).r_n
\end{equation}
where $\varepsilon_s$ is the channel constant, $r_s$ denotes the symbol rate per seconds, $M$ is the number of channels, and $r_n$ is the total capacity of the system given by $(M\times S)$ while the  probability $F_i$ of the PU channel being busy or idle is given as: 
\[ F_i=
\Big\{
\begin{tabular}{cc}
 1 & $i  \in [1,M]$ \\
 0 & $\textit{otherwise}$ 
\end{tabular}\]
Note that  $\vartheta_i=A_i/(A_i+C_i)$  is the channel utilization ratio. Therefore, the   primary user’s channel slot capacity $\varphi_P$, is given as \cite{esenogho2015b}:
\begin{equation}
\varphi_P=S.\sum_{i=1}^{M}\frac{A_i}{A_i+C_i}=S.\sum_{i=1}^{M}(\vartheta_i).
\end{equation}
While SU channel slot capacity $\theta_{su}$ is the reminder after PU occupancy. It can be expressed as:
\begin{equation}
\theta_{su}=(M\times S)-\varphi_P
\end{equation}
\vspace{7pt}
\section{Channel Assembling Strategies}\label{sec4}
\subsection{Instant blocking strategy with queue $(IBS+Q)$ $(\theta_{pu}, \theta_{su}, \theta_{i}, \theta_{j}, q)$}
In $IBS+Q$ scheme, if there are no resources upon SU request the SU instead of Instantly blocking it request, it is queue in $q$ though can be block and dropped if and only if the queue is full and the SU over stayed on the queue.  However, a SU that arrives first is given higher priority to the SU that arrives last and all the SUs are pre-empted by the arrival of a   primary user \cite{esenogho2015b}. 
Assuming that, $K$ denotes the number of SUs on the network and the resource demanded by $SU_i$, $SU_j$ are $\theta_{i}$  and $\theta_{j}$. The aggregation protocol is shown in Table I.
\begin{table}[h]\label{tb1}
	\centering
	\caption{Algorithm for $IBS+Q$ scheme}
\begin{tabular}{l}
	\hline\hline
	\textit{Algorithm for} $IBS+Q$ \textit{scheme}\\
	\hline
	\textbf{\textit{CRBS check wireless link;}} // \textit{cognitive radio base station checks}\\ 
	\textit{wireless link state (SNR) for SUs.}\\
	\textbf{\textit{CRBS check} $\theta_{su}$}; //\textit{cognitive radio base station checks}\\
	\textit{available recourse for SU}\\
	\textbf{\textit{If}} $(\theta_{su}\ge \sum_{i=1}^{K}\theta_i);$ // \textit{test for  $SU_i$ resources}\\
	$\mbox{ }\mbox{ } SU_{i,j}\_admit=true$; // \textit{admit $SU_i$ and aggregate resources for it}\\
	\textbf{\textit{Else}}\\
	$\mbox{ }\mbox{ } SU_{i}\_admit=false$; // \textit{block the $SUs$ due to insufficient resources}\\
	\textbf{\textit{If}} $(q_{1}< q_{1max});$ // \textit{queue not full to the max}.\\
	$\mbox{ }\mbox{ } SU_{i,j}\_admit\_queue=true$; // \textit{allow new $SU_i$}\\
	\textbf{\textit{Else}}\\
	$\mbox{ }\mbox{ } SU_{i,j}\_admit=false;$ // \textit{queue-full block $SU_i$}\\
	\textbf{\textit{Else}}\\
	\textbf{\textit{If}} ($\delta_{SUi,j}>\delta_{max}$); // \textit{over delay in the buffer}\\
	$\mbox{ }\mbox{ } SU_{i,j}\_drop\_queue=true$ \textit{(forced terminate)}; // \textit{time-out in the queue}\\
	\textbf{\textit{End if}}\\
	\textbf{\textit{Go to start}}\\
	\hline\hline
\end{tabular}
\end{table}
\subsection{$RBS+Q$ $(\theta_{pu}, \theta_{su}, \theta_{i,j}^{min}, \theta_{i,j}^{max}, q)$}
In this scheme, both  $SU_i$ and $SU_j$ requires a minimum and maximum of  $\theta_{i,j}^{max}$ and $\theta_{i,j}^{min}$ number of channels to commence or stop assembling respectively. If a $SU_i$ require service, the aggregation procedure checks for resource availability similar to the $IBS+Q$ scheme. If the resources are available and sufficient, $SU_i$ is admitted, otherwise, readjustment algorithm is executed to reduce blocking and dropping. Every other procedure remains the same. The aggregation protocol is shown in Table II.
\begin{table}[h]\label{tb2}
	\centering
	\caption{Algorithm for $RBS+Q$ scheme}
	\begin{tabular}{l}
		\hline\hline
		\textit{Algorithm for $RBS+Q$ scheme}\\
		\hline
		\textbf{\textit{CRBS check wireless link} $\gamma_{su}$;} // \textit{cognitive radio base station checks}\\ 
		\textit{wireless link state (SNR) for SUs.}\\
		\textbf{\textit{CRBS check $\theta_{su}$;}} //\textit{cognitive radio base station checks}\\
		\textit{available recourse for SU}\\
		\textbf{\textit{If}} $(\theta_{su}\ge \sum_{i=1}^{K}\theta_{n_i}^{min});$ // \textit{test for  $SU_i$ resources}\\
		$\mbox{ }\mbox{ } SU_{i,j}\_admit=true$; // \textit{admit $SU_i$ and aggregate}\\
		\textbf{\textit{Else}}\\
		$\mbox{ }\mbox{ }$ \textbf{\textit{Do}} $\theta_n^{max}-1$, $++SUs;$ // \textit{SUs with highest channel donate}\\
		\textit{ to new SUs and iterate over other higher SUs resources.}\\
		\textbf{\textit{Else}}\\
		 \textbf{\textit{If}} $(q_{1}< q_{1max});$ // \textit{queue not full/empty}.\\
		$\mbox{ }\mbox{ } SU_{i,j}\_admit=true$; // \textit{allow SU into buffer to wait}\\
		\textbf{\textit{Else}}\\
		$\mbox{ }\mbox{ } SU_{i,j}\_admit=false$; // \textit{queue-full block $SU_i$}\\
		\textbf{\textit{Else}}\\
		\textbf{\textit{If}} $(\delta_{SUi,j}>\delta_{max});$ // \textit{over delay in the buffer}\\
		$\mbox{ }\mbox{ } SU_{i,j}\_drop\_queue=true$ \textit{(forced terminate)}; // \textit{timeout in the queue}\\
		\textbf{\textit{End if}}\\
		\textbf{\textit{Go to start}}\\
		\hline\hline
	\end{tabular}
\end{table}
\vspace{7pt}
\section{System model performance analysis}\label{sec5}
The performance analysis is based on the premise of \cite{esenogho2015b}. Arrival rates of the PU and SU follows a \textit{Poisson distribution} while the service time is exponentially distributed with the service and arrival rates well-defined as $\mu_p$, $\mu_s$, $\lambda_p$, and $\lambda_s$, for the PU and SU respectively. The total rate for a SU is taken as the multiplication of the SU channel service rate and the number of aggregated channels $\theta_i\mu_{si}$ or $\theta_j\mu_{sj}$.
\begin{enumerate}
	\item 	\textit{Blocking probability} $(P_b)$  of SU is the fraction of total SU blocked to the total SU arrived. If total SU blocked  and arrived are  $\Omega_{tsu}$ and $\lambda_{tsu}$ respectively, then , It is expressed as:
	\begin{equation}
	\begin{split}
	P_b&=\frac{\textit{Sum of secondary user blocking rate}}{\textit{Sum secondaty user arrival rate}}\\
	&=\frac{\Omega_{tsu}}{\lambda_{tsu}}
	\end{split}
	\end{equation} 
	\item 	\textit{The forced termination probability} $(P_f)$, ratio of total SU dropped to the total admitted SU connection. Similarly, If total SU forced terminated and admitted are  $\eta_{tsu}$ and $\zeta_{tsu}$ respectively.it is expressed as:
	\begin{equation}
	\begin{split}
	P_f=&\frac{\textit{Total secondary user forced terminated}}{\textit{Total admitted SU connections}}\\
	=&\frac{\eta_{tsu}}{\zeta_{tsu}}
	\end{split}
	\end{equation}
	\item \textit{Access Probability} $(P_a)$, is defined as the likelihood that enough resources exist for the SU when it arrives after meeting the necessary conditions in the algorithms. It can be given as:
	\begin{equation}
	P_a=1-P_b
	\end{equation}
	\item \textit{Queue size} $\delta_l$: the length of the queue depends of the arrival rates of the SUs and service completion rates.\\
	
	\item \textit{Capacity} $\rho_{su}$: the capacity of the SU traffic is the mean number of SU’ service completion per unit time.  Thus, $\rho_{su}$ of accepted SU at a particular time is dependent on the SNR per modes pair. It is expressed as:
	\begin{equation}
	\rho_{su}=\frac{\textit{mean number of SUs service completion}}{\textit{Time (seconds)}}
	\end{equation}
\end{enumerate} 
\vspace{7pt}
\section{Numerical results and discussions}\label{sec6} 
\begin{figure}[h]
	\centering
	\includegraphics[width=1\linewidth]{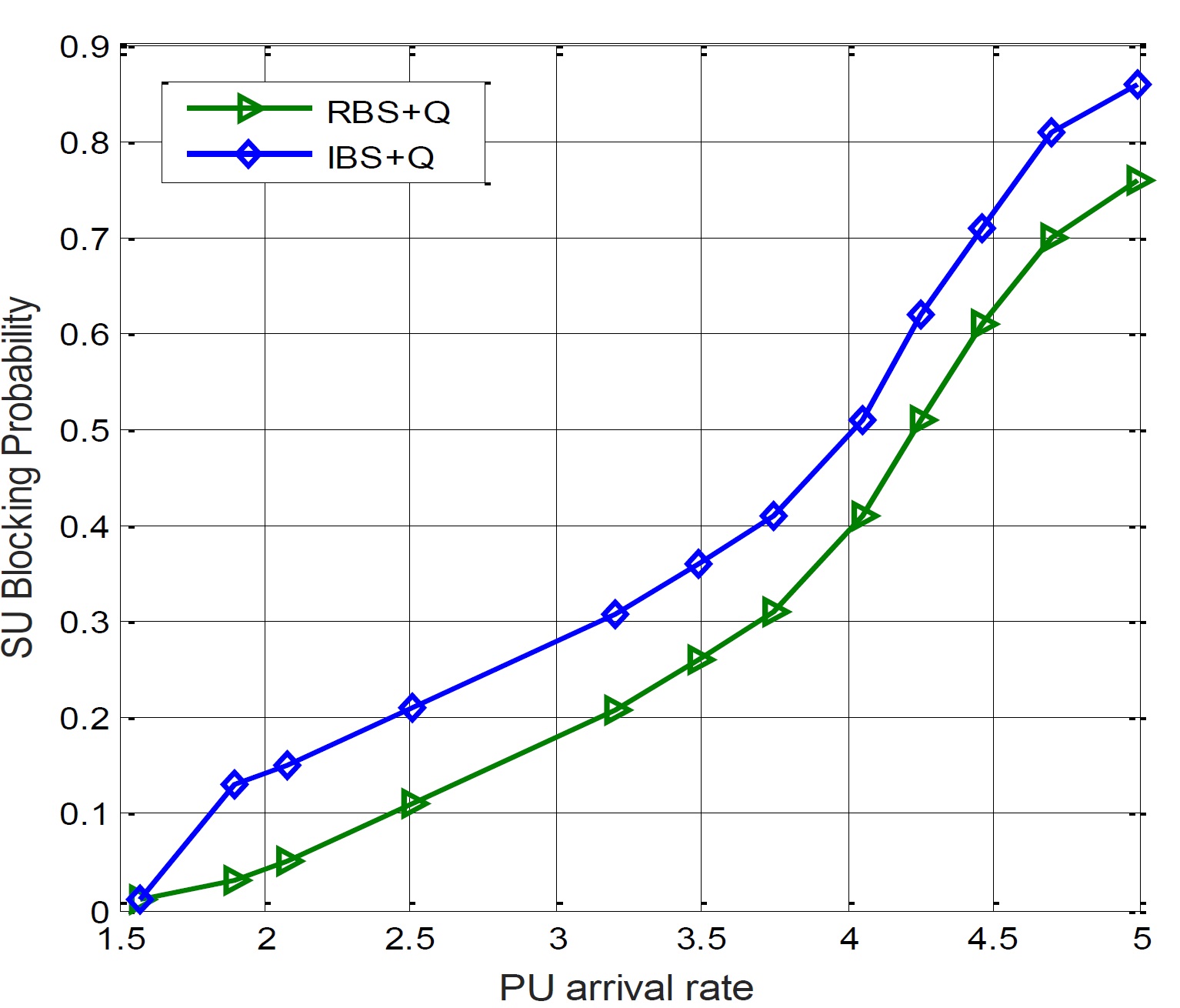}
	\caption{$P_b$ vs. $\lambda_p$.}
	\label{fig:fg4}
\end{figure}

This section presents the performance of the two CA policies with queueing regime. The numerical results are based on the system simulation. Parameters are set as in \cite{esenogho2015b}. 
In Fig. \ref{fig:fg4} the increase in $P_b$ is a function of PU arrival $\lambda_p$. This growth is due to batch arrival of PU into their spectrum. This however, impact the SU' service since most channel slots are occupied by the PU and as such, access will be deprived to SUs. Therefore, the blocking probability increases. However, in this case, the $RBS+Q$ results outperformed the $IBS+Q$ due to its flexibility.

\begin{figure}[h]
	\centering
	\includegraphics[width=1\linewidth]{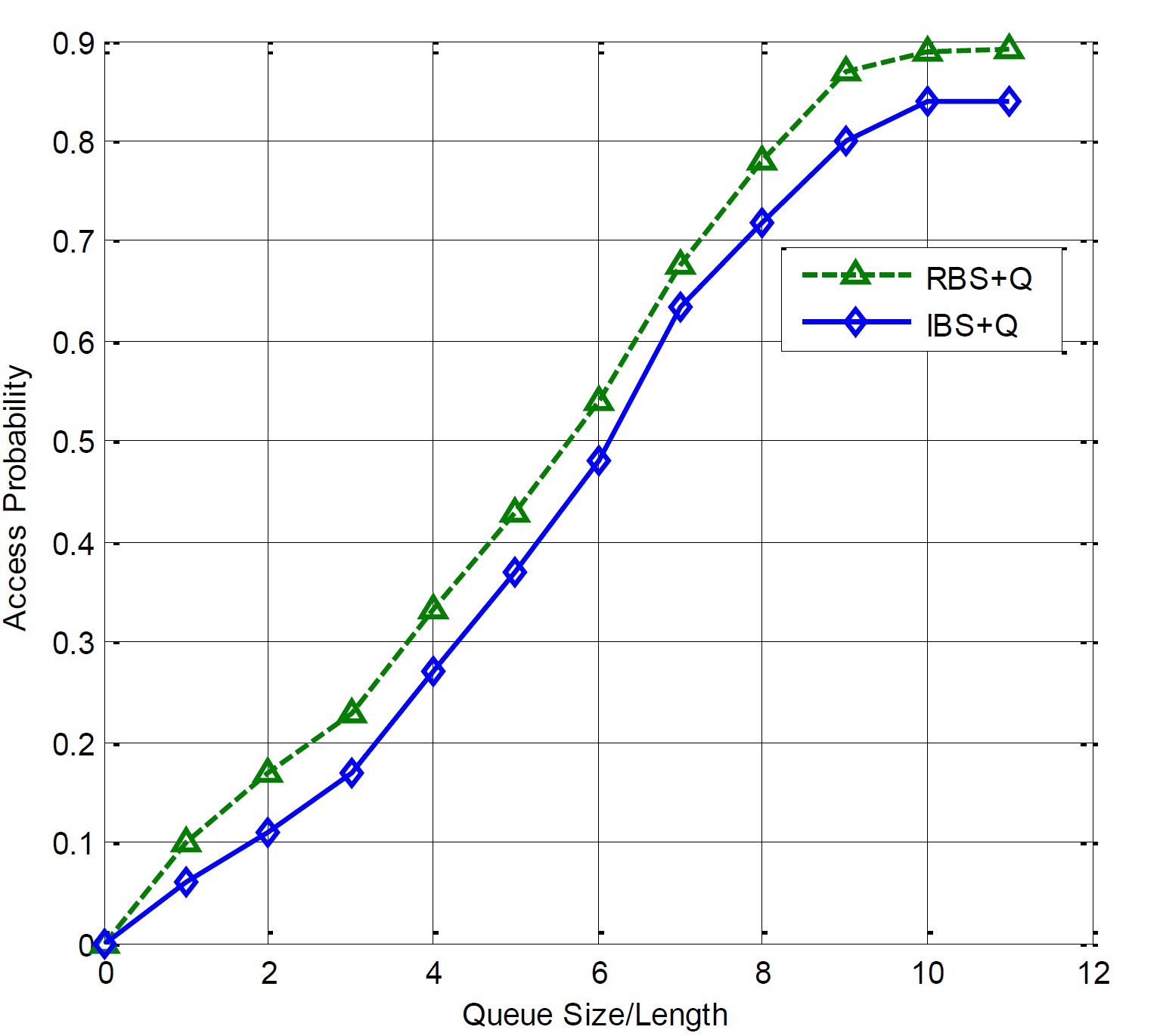}
	\caption{$P_a$ vs. $\delta_l$.}
	\label{fig:fg5}
\end{figure}

\begin{figure}[h]
	\centering
	\includegraphics[width=1\linewidth]{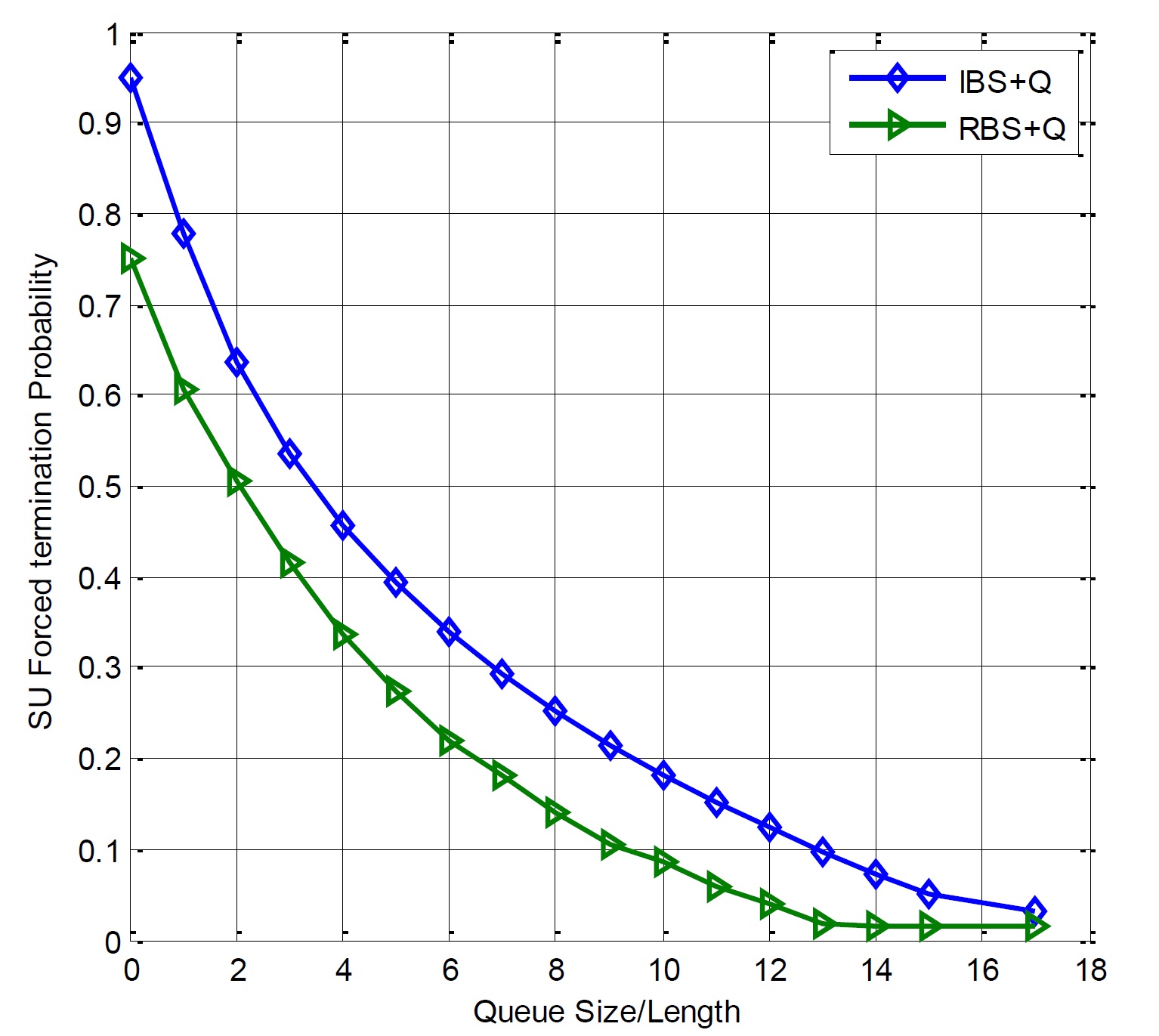}
	\caption{$P_f$ vs. Queue size.}
	\label{fig:fg6}
\end{figure}

Fig. \ref{fig:fg5} show the effect of incorporating buffer regime into CA. This gives the SUs the opportunity (avenue to wait) to access the spectrum whenever the PU interrupts their services or if the SU experiences insufficient or no resources.  As the queue length increases, the likelihood of SU accessing the scarce resources grows and at a point begins to saturate due limited buffer capacity. For the sake of space, more results would have been presented but will be considered in future work particularly the response of access probability on PU arrival rate, which normally is expected to decrease. 
Fig. \ref{fig:fg6} explains the impact of the queuing regime on the forced termination probability as soon as the PU interrupts SU traffic flow. As the queue size rises (more SUs being buffered) the forced termination probability drops significantly since a "second chance" will be given to SUs. However, when a SU exceeds its waiting time in the queue irrespective of the class, it will be dropped. This is to avoid starvation of other SUs waiting in the queue.
\balance

\begin{figure}[h]
	\centering
	\includegraphics[width=1\linewidth]{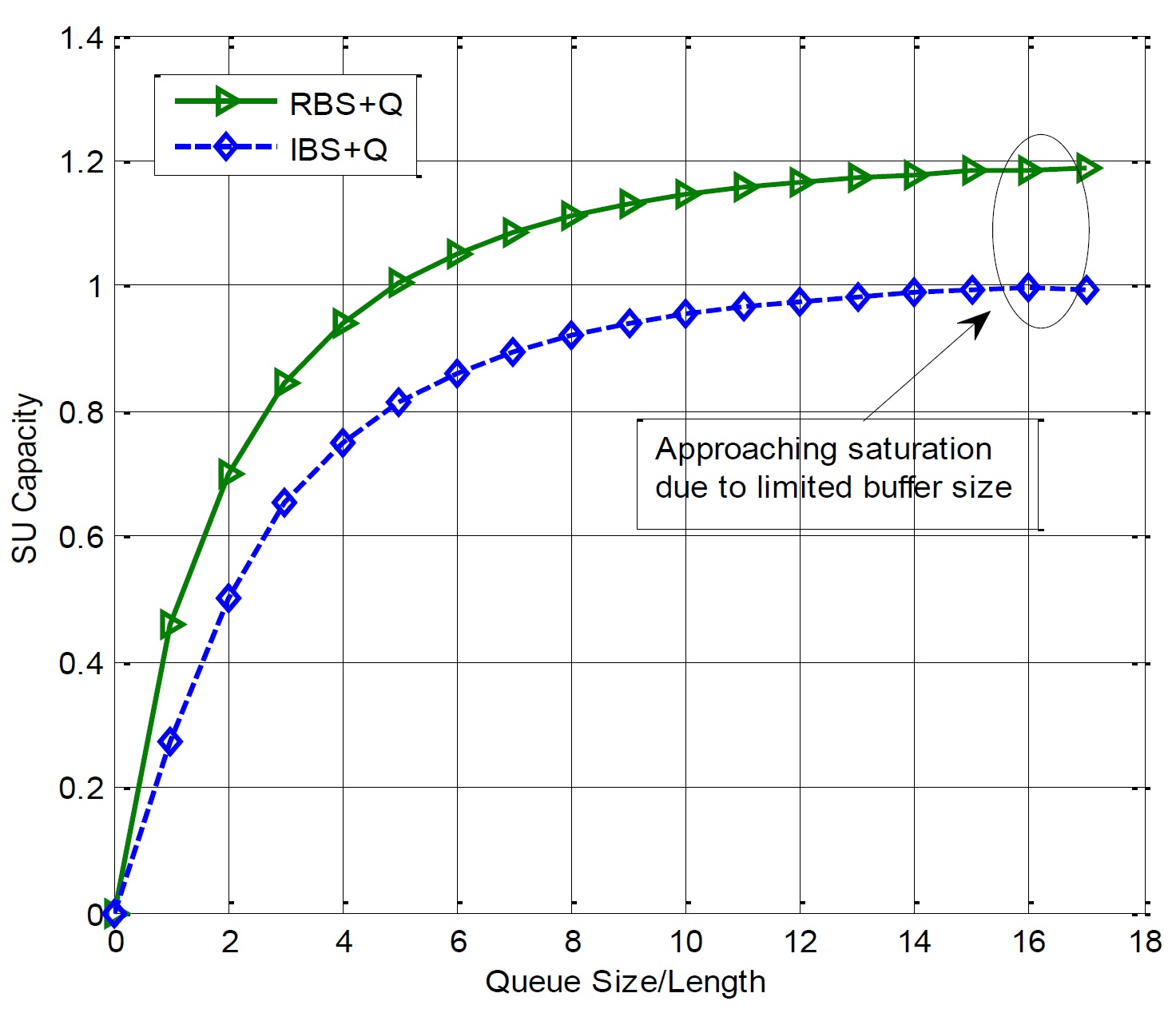}
	\caption{$SU$ Capacity vs. Queue size.}
	\label{fig:fg7}
\end{figure}

In Fig. \ref{fig:fg7}, both policies showed improved SUs capacity as the queue size increases. This implies that more SUs have been given the opportunity to transmit their packets that would have been dropped when the PU arrives while SUS are still using the spectrum. However, the flexibility of the $RBS+Q$ is still an advantage over the $IBS+Q$.
\vspace{7pt}
\section{Conclusion}\label{sec:conclusion}
By comparing the two strategies $RBS+Q$ and $IBS+Q$, there is a substantial superiority of the $RBS+Q$ scheme over the $IBS+Q$ due to its adaptability. Precisely, the $RBS+Q$ scheme outperformed the $IBS+Q$ scheme in terms of SU blocking, forced termination, and access probabilities respectively. It demonstrates that $AMC$ with queueing technique is a robust method in improving CA schemes. Our future work will include a detailed performance analysis using continuous time Markov chain $(CTMC)$ and other queuing regimes. 
\vspace{7pt}
\section*{ACKNOWLEDGMENT}
This work is supported by Prof Ferreira's Centre for telecommunication (CfT) under the Global Excellence Stature Post-Doctoral Research Fellowship program at the University of Johannesburg, South Africa.


\vspace{7pt}

\begin{thebibliography}{11}
\bibitem{liang2011}
Y. Liang, K. Cheng, G. Ye. Li, and P. Mahonen, ``Cognitive radio network and communication: an overview,'' \textit{IEEE Transactions on Vehicular Technology}, vol. 60, no. 7, pp. 3386-3407, Sept. 2011. 
\bibitem{domenico2012}
A. Domenico, E. C. Strinati, and M. G. Benedetto, ``Survey on MAC strategies for cognitive radio network,'' \textit{IEEE Communication Survey and Tutorials}, vol. 14, no. 1, pp. 21-43, 2012. 
\bibitem{nguyen2012}
V. T. Nguyen, F. Villain, and Y. L. Guillou, ``Cognitive radio RF: overview and challenges,'' VLSI Design, vol. 2012, pp. 1-13, Feb. 2012. 
\bibitem{su2008}
H. Su and X. Zhang, ``Cross-layer based opportunistic MAC protocol for QoS providing over cognitive radio wireless networks,'' \textit{IEEE Journal on Selected Area in Communication}, vol. 26, no. 1, pp. 118-129, Jan. 2008. 
\bibitem{jiao2012}
L. Jiao, F. Y. Li, and V. Pla, ``Modelling and performance analysis of channel assembling in multi-channel cognitive radio networks with spectrum adaptation,'' \textit{ IEEE Transactions on Vehicular Technology}, vol. 61, no. 6, p. 2686–2697, Jul. 2012. 
\bibitem{jiao2013}
L. Jiao, E. Song, V. Pla, and F. Y. Li, ``Capacity upper bound of channel assembling in cognitive radio networks with quasi-stationary  primary user activities,'' \textit{IEEE Transactions on Vehicular Technology}, vol. 62, no. 4, pp. 1849-1855, May 2013.
\bibitem{balapuwaduge2014}
I. A. M. Balapuwaduge, L. Jiao, V. Pla, and F. Y. Li, ``Channel assembling with priority-based queue in cognitive radio network: strategy and performance,'' \textit{IEEE Transaction on wireless Communication}, vol. 13, no. 2, pp. 630-645, Feb. 2014. 
\bibitem{esenogho2015}
 E. Esonogho and T. Walingo, ``Performance evaluation of channel assembling strategies with multi-class secondary users in cognitive radio networks,'' \textit{Southern Africa Telecommunication, Networking and Application Conference}, Cape Town, South Africa, Sept. 6-9,  2015, pp. 81-86. 
\bibitem{esenogho2015b}
E. Esenogho and T. Walingo, ``evaluation of channel assembling strategies with single-class secondary users in cognitive radio networks,'' \textit{IEEE Wireless Vitae}, Hyderabad, India, Dec. 13-16, 2015. 
\bibitem{ngatched2013}
T. M. N. Ngatched, S. Dong, and A. S. Alfa, ``Analysis of cognitive radio networks with channel assembling, buffering, and imperfect sensing,''  \textit{IEEE Wireless Communications and Networking Conference}, Shanghai, China, Apr. 7-10  2013, pp. 952-957. 
\bibitem{li2012}
L. Li, S. Zhang, K. Wang, and W. Zhou, ``Queuing method in combined channel assembling and fragmentation strategy for dynamic spectrum access,'' \textit{IEEE Symposium on personal, Indoor, and Mobile Radio Communication}, Sydney, Australia, Sept. 9-12, 2012, pp. 1214-1219.
\bibitem{jiao2011} 
L. Jiao, F. Y. Li, and V. Pla, ``Greedy versus dynamic channel assembling strategy in CRNs: Markov models and performance evaluation,'' \textit{International Networking Conference}, Valencia, Spain,  May 9-13, 2011, pp. 22-31. 
\bibitem{jiao2010}
L. Jiao, F. Y. Li, and V. Pla, ``Analysis on channel bonding/assembling for multi-channel cognitive radio networks,''  \textit{IEEE European Wireless Conference}, Lucca, Italy, Apr. 12-15, 2010, pp. 468-474.
\bibitem{wang2007}
X. Wang, Q. Liu, and G. B. Giannakis, ``Analyzing and optimizing adaptive modulation coding jointly with ARQ for QoS-guaranteed Traffic,'' \textit{IEEE Transactions On Vehicular Technology}, vol. 56, no. 2, pp. 710-720, Mar. 2007. 
\bibitem{razavilar2002}
J. Razavilar, K. J. R. Liu, and S. I. Marcus,, ``Jointly optimized bit-rate/delay control policy for wireless packet networks with fading channels,'' \textit{IEEE Transactions on Communication}, vol. 50, no. 3, pp. 484–494, Mar. 2002. 
\end{thebibliography}
\end{document}